\documentclass[prl,twocolumn,superscriptaddress]{revtex4}

\usepackage[dvips]{graphicx}
\usepackage{color}
\usepackage{latexsym}
\usepackage{amsmath}
\usepackage{amsthm}
\usepackage{amssymb}
\usepackage{amsfonts}

\begin{document}
\title{Finite size Berezinski-Kosterlitz-Thouless transition
at grain boundaries in solid $^4$He and role of $^3$He impurities}

\author{Sergio~Gaudio}
\affiliation{Dipartimento di Fisica, Universit\`a ``La Sapienza'',
P.le A. Moro 2, 00185 Rome, Italy}
\affiliation{SMC Research Center and ISC, INFM-CNR, v. dei Taurini 19,
00185 Rome, Italy}
\author{Emmanuele~Cappelluti}
\affiliation{Dipartimento di Fisica, Universit\`a ``La Sapienza'',
P.le A. Moro 2, 00185 Rome, Italy}
\affiliation{SMC Research Center and ISC, INFM-CNR, v. dei Taurini 19,
00185 Rome, Italy}
\author{Gianluca~Rastelli}
\affiliation{ Laboratoire de Physique et Mod{\'e}lisation 
des Milieux Condens{\'e}s,   Universit{\'e} Joseph Fourier, \\
CNRS - UMR 5493, BP 166, 38042 Grenoble, France}
\author{Luciano~Pietronero}
\affiliation{Dipartimento di Fisica, Universit\`a ``La Sapienza'',
P.le A. Moro 2, 00185 Rome, Italy}
\affiliation{SMC Research Center and ISC, INFM-CNR, v. dei Taurini 19,
00185 Rome, Italy}
\date{\today}

\begin{abstract}
We analyze the 
complex phenomenology of the Non-Classical Rotational Inertia (NCRI)
observed at low temperature in solid $^4$He
within the context 
of a two dimensional Berezinski-Kosterlitz-Thouless transition
in a premelted $^4$He film at the grain boundaries.
We show that both the temperature and $^3$He doping dependence
of the NCRI fraction (NCRIF)
can be ascribed to finite size effects induced by the
finite grain size. 
We give an estimate of the average size
of the grains which we argue to be limited by the isotopic
$^3$He  impurities and we provide a simple power-law relation between
the NCRIF and the $^3$He concentration.
\end{abstract}
\maketitle

The report of a Non Classical Rotational Inertia (NCRI)
in solid $^4$He \cite{kc1,kc2}
has opened an intense debate
in the physics community
about its possible ``supersolids'' (SS) nature.
Although the observation of NCRI
has been confirmed by other groups \cite{jap,ann,aoki},
its phenomenology presents strong discrepancies with a
simple supersolid phase, so that
the precise origin of this phenomenon is still unclear.
On one hand the supposed SS transition
appears to be anomalously broad in temperature \cite{kc1,kc2,jap,ann,aoki}.
On the other hand the NCRI
strongly depends on the
external conditions \cite{ann,aoki,RR,kc3,kc4}.
In Refs. \cite{ann,RR} for instance annealing was shown to
reduce and even to make disappear the NCRI fraction (NCRIF).
Moreover,
using a different set-up, in Ref. \cite{balibar}
it was shown that the mass flow, associated with a SS phase,
occurred only in the presence of
grain boundaries (GB), and it was absent when GBs were
not detected.
This observation gives rise then to
an alternative hypothesis to the SS phase, namely, that a liquid phase
is confined at the GBs and that the mass flow
is related to superfluidity of the liquid component,
similarly to a Rollin film.
Partial wetting of GBs was experimentally observed in Ref. \cite{sasaki2},
and the possible superfluid (SF) ordering was investigated
in Refs. \cite{PS,pollet,rossi}.
Interestingly enough, a change of the shear modulus has also been
observed at the low temperature, with a similar dependence
on annealing and on $^3$He concentration as the NCRIF \cite{day0}.
The connection between these two quantities is thus worth
to be further investigated.

Quite puzzling is also the dependence of the NCRI phenomenology
on the $^3$He concentration $x_3$.
The first report of NCRI \cite{kc1} was observed in commercial $^4$He,
which contains generally a low concentration
$x_3 \sim 0.3$ ppm.
Further investigations
showed that the critical temperature increases monotonically
with $x_3$ whereas NCRIF increases with $x_3$ only up to
an optimal doping at $x_3 \sim 300$ ppb, after which
the magnitude decreases \cite{chanHetre}.

In this paper we propose that,
due to the strong confinement on the grain boundaries,
the SF transition of the premelted liquid component
can be described in terms of a
two dimensional Berezinski-Kosterlitz-Thouless (BKT)
superfluid transition where the grain size
gives rise to finite size effects.
We propose also a simple model where
the concentration of $^3$He impurities rules the
grain size, and hence the finite size effects.
We show that this framework can explain in a natural way,
for $x_3 \le 300$ ppb,
the broadness of the SF transition 
and the dependence of the NCRIF on the
$^3$He impurity concentration.

In the following we shall model the polycrystal $^4$He samples
in terms of spherical grains with radius $R$
and probability distribution function $P(R)$.
Premelting effects, as discussed in Refs. \cite{sasaki2,PS,pollet,rossi,SM},
is expected to give rise to a thin liquid film with thickness $d$.
Partial wetting \cite{sasaki2,pollet},
reducing the liquid amount covering the grain,
can be also considered but it will not change our results.
We assume that $d \ll R$, so that the liquid helium system
confined on the GB surface can be regarded as two-dimensional.
This kind of model was employed by Kotsubo and Williams (KW)
to explain the behavior of SF $^4$He films on different
substrates \cite{KW}.
The important ingredient within this context is that the size $R$
of the grain provides an intrinsic finite size cut-off which
makes the BKT transition to be smooth. The broadness of the NCRI
transition can be thus employed to estimate the average
size $R_0$ of the GBs.

The BKT self-consistent equations on a spherical geometry were discussed
in Ref. \cite{KW} by KW.
We can define the energy $U_0(\theta)$ of an isolated vortex-antivortex pair at
angular distance $\theta$, in units of $k_{\rm B}T$, as
\begin{equation}
U_0(\theta)
=2U_c+
\int_{2\theta_c}^\theta
\frac{\pi K_0}{\tan\left[(\theta'-\theta_c)/2\right]}
d\theta',
\label{U_0}
\end{equation}
where 
$K_0 = \hbar^2 \sigma^0_s/m^2k_{\rm B}T$ is related
to the bare areal SF density $\sigma^0_s$, $m$ is the
$^4$He mass, $U_c$ is the vortex core energy, and
$\theta_c=a_0/R$ is the minimum vortex-antivortex angular distance
given by the vortex core size $a_0$.
In the presence of screening effects
due to vortex pair polarization,
we can generalize Eq. (\ref{U_0}) as
\begin{equation}
U(\theta)
=2U_c+
\int_{2\theta_c}^\theta
\frac{\pi K_0}{\epsilon_0(\theta)
\tan\left[(\theta'-\theta_c)/2\right]}
d\theta',
\label{U}
\end{equation}
where the static dielectric constant $\epsilon_0(\theta)$
can be evaluated in a self-consistent way as:
\begin{equation}
\epsilon_0(\theta)=
1+\frac{4\pi^3K_0}{(2\theta_c)^4}
\int_{2\theta_c}^\theta d\theta'
\theta'^2\sin\theta'\exp[-U(\theta')].
\label{die}
\end{equation}

Eqs. (\ref{U_0})-(\ref{die}) can
be evaluated self-consistently for all $\theta \le \pi$
to obtain the observable SF density
$\sigma_s(T)=\sigma_s^0/\epsilon(\theta=\pi)$
as function of temperature.
Eqs. (\ref{U_0})-(\ref{die}) can be also generalized to
the dynamical case by introducing
the dynamic dielectric constant $\epsilon(\theta,\omega)$
which depends on the
diffusion constant of the vortices $D$ and on the angular frequency $\omega$
through the parameter $r_D = \sqrt{2 D/\omega}$ \cite{KW,AHNS,WY}.
In the physical range of the experimental setup $r_D/a_0 \gg 1$
and the evaluation of the SF density $\sigma_s$ in the
dynamical regime is practically indistinguishable from the static one.
The introduction of the dynamical analysis permits however the evaluation
as well of the change of quality factor 
$\Delta[Q^{-1}]$ \cite{KW,AHNS,WY}.

Let us now apply the above analysis to our polycrystal
spherical-grain model.
For a single grain of size $R$, assuming $d \ll R$,
we can estimate the temperature dependent  NCRIF $n_s(T,R)$ as
\begin{equation}
n_s(T,R) \simeq \frac{4\pi R^2 \sigma_s(T)}{4\pi \rho R^3/3}
= \frac{3\sigma_s}{\rho R},
\label{MCMI}
\end{equation}
(where $\rho$ is the solid $^4$He density),
and a  mean $n_s(T)=\int dR P(R) n_s(T,R)$.
In the following we shall show that the SF temperature profile
is mainly ruled by the mean grain size value $R_0$.
In this case we can roughly estimate the zero temperature NCRIF
$n_s \simeq 3\sigma_s^0/\rho R_0$.
Note also that in the $R_0\gg a_0$ limit the areal
SF density $\sigma_s^0$ is roughly proportional to
the Berezinski-Kosterlitz-Thouless temperature $T_{\rm BKT}$
where $\sigma_s(T)$ drops to zero. Within this context thus the effect
of finite GB size $R$ is mainly to give rise to a significant broadness
of the transition whereas the temperature position of the drop
is only weakly affected. If we assume
$\sigma_s^0\simeq \sigma_s(T_{\rm BKT})=(2m^2/\pi\hbar^2)k_{\rm B}T_{\rm BKT}$,
and we estimate $T_{\rm BKT}$ from the temperature $T_{50}$ at which
$n_s(T)$ drops to its 50 \% value of $n_s(T=0)$,
we can thus obtain a free fitting parameter estimate
of the GB size:
\begin{equation}
R_0 \simeq \frac{6m^2}{\pi \rho\hbar^2}\frac{k_{\rm B}T_{\rm BKT}}{n_s}.
\label{rmodel}
\end{equation}
We would like to stress that, because of the simplicity of
this model and of the slight approximations in the estimates
of $T_{\rm BKT}$ and $\sigma_s^0$, Eq. (\ref{rmodel}) is simply meant
to give the order of magnitude of $R_0$.

We now apply our model to the specific case of NCRI in solid
$^4$He. We consider a $^4$He density $\rho = 0.2$ g/cm$^3$,
which corresponds to a molar volume $\sim 20$ cm$^3$ and to a pressure
$41$ bar. We set also typical values for the vortex core size
$a_0 = 50$ \AA\ 
and energy $U_c=2.5K_0$ \cite{cho}.
We consider for the moment two extreme probability distribution functions,
namely a single value $P(R)=\delta(R-R_0)$ and a flat
$P(R)=1/R_0$ for 
$R_0-\Delta R_0 \le R \le R_0+\Delta R_0$, with
$\Delta R_0=R_0/2$.
With this choice of parameters, the overall profile of
the NCRIF $n_s(T)$ is uniquely
determined by the only two free parameters,
namely $R_0$ and $\sigma_s^0$,
where $\sigma_s^0/R_0$ rules the magnitude of $n_s$ at $T=0$ while
$R_0$ is related to the broadness of the SF transition.

\begin{figure}[t]
\includegraphics[width=2.65in]{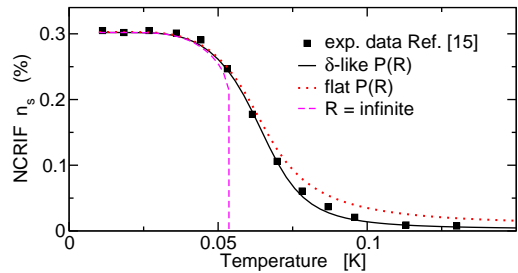}
\caption{(Color online) 
Comparison between experimental data for the 133 ppb TOP sample
(symbols) \cite{chanHetre}
and our theoretical analysis. Also shown is the BKT curve for $R=\infty$.
}
\label{f-133ppb}
\end{figure}

In order to show the feasibility of this approach to reproduce
the experimental results, we compare in Fig. \ref{f-133ppb}
the $n_s$ vs. $T$ data for the 133 ppb $^3$He Penn State (TOP)
sample \cite{chanHetre}
with our best fit, which gives $\sigma_s^0=0.26 \times 10^{-9}$ g/cm$^2$
and $R_0=130$ \AA,
and with a infinite size BKT transition for large grains $R \gg a_0$.
Note that there is only a slight difference between
the $\delta$-like and the flat $P(R)$. As a matter of fact,
we have checked that different distribution functions $P(R)$
do not affect qualitatively our results, so that from now on
we shall consider for simplicity a simple $\delta$-like
distribution.
The nice agreement between the experimental data and our results
suggests that the NCRI broad transition is not related to
inhomogeneities of the samples but it stems from finite size
effects due to the finiteness of the $^4$He grain.
It is important to underline that, while the fitting procedure
gives a refinement of the GB size, the order of magnitude of $R$
is essentially given by the relation (\ref{rmodel}), so that the
good agreement in the broadness of the transition is not
a result of the fit but it can be considered as an independent check
of the validity of our analysis.

The estimate size $R_0 \simeq 130$ \AA\ of the grains can appear quite
puzzling especially considering that $^4$He
at these conditions is thought to solidify
in a polycrystal form with macroscopic size of grains.
However, on one hand the crystallographic evidence cannot
esclude mosaics of small-angle grain boundaries.
On the other hand, the presence of such small grains
can naturally account for the high sensitivity to annealing
and to preparation and freezing procedures, and the similarities
with the NCRI phenomenology in Vycor glasses \cite{ceperley}.
Nevertheless, the physical origin of such an extremely small grain size
and its dependence on $^3$He concentration requires to be explained.
We address this issue in the second part of this Letter, where
we relate the origin of GB to the presence of liquid melted bubbles
induced by $^3$He impurities.

In order to gain a further insight on this point, let us discuss before
the experimental dependence of the NCRI phenomenology
on the He$^3$ concentration and
on the growth/measurement condition.
In particular, we analyze two samples \cite{chanHetre}
in the highly dilute limit $x_3 \simeq 1$ ppb, one belonging to
the University of Florida (TOF) family, grown with the blocked capillary
method, and one still belonging to the Penn State University but
grown at constant pressure (CP).
For both these samples measurements of
the change of the quality factor $\Delta[Q^{-1}]$ were
also available \cite{chanHetre}.

\begin{figure}[t]
\includegraphics[width=3.3in]{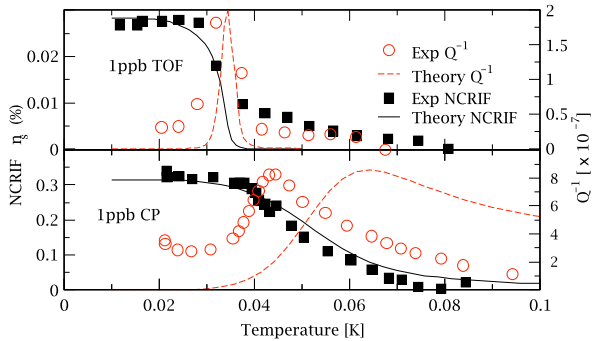}
\caption{(Color online) Comparison between theory and experimental NCRIF and
$\Delta[Q^{-1}]$ data taken from Ref. \cite{chanHetre}
for the 1 ppb TOF (upper) and CP (lower) samples.}
\label{f-compare}
\end{figure}

The first thing to observe is that, despite both samples are at same $x_3$,
the signals are significantly different (Fig. \ref{f-compare}).
For the TOF sample, the value
of $n_s$ is almost one order of magnitude smaller than for
the CP sample. Its variation as a function of the temperature
is also different. In fact, while 
the $n_s$ presents a quite smooth behavior in the CP sample,
the drop of $n_s$ at $T \simeq 0.03$ K
is much sharper followed by an additional tail.
Similar features are observed in the behaviors of 
the quality factor $\Delta[Q^{-1}]$ which is
quite broad in the CP sample compared to the sharp peak
at $T \simeq 0.03$ K in the TOF sample
(note that the longer tail in the $n_s$
is not observed in $\Delta[Q^{-1}]$,
suggesting that this features in the $n_s(T)$ behavior could
have a spurious origin).

We can see now that our model permits to understand in a very natural way
also the $x_3$ dependence of the NCRI phenomenology.
We first note that, although the two samples have different $n_s(T=0)$
of a order of magnitude, their $T_{\rm BKT}$, defined for instance
as $T_{\rm BKT}=T_{50}$, are of a similar magnitude.
A simple analysis, using again Eq. (\ref{rmodel}), would point out thus
an average radius $R_0$ of the GBs in the TOF sample one order
of magnitude larger than in the CP one.
As a consequence, finite size effects are expected to give rise to a much
sharper transition in the TOF than in the CP case, in agreement
with the experimental observation. These simple considerations
are corroborated
by a more detailed analysis.
In Fig. \ref{f-compare} we show our best fits
for both the TOF and CP samples,
compare with the experimental data.
Estimates for $R_0$ and $\sigma_0$, in these cases, are
respectively
$R_0=790$ \AA, $\sigma_s^0=0.105 \times 10^{-9}$ g/cm$^2$,
and $R_0=100$ \AA, $\sigma_s^0=0.216 \times 10^{-9}$ g/cm$^2$.
For the evaluation of the
quality factor $\Delta[Q^{-1}]$ we have used respectively
$r_D=1.2\times 10^4$ \AA\ and $r_D=0.3\times 10^4$ \AA,
which give $r_D/R \gg 1$.
We remind that in such quasi-static regime the parameter
$r_D$ simply acts as a scale factor on
$\Delta[Q^{-1}]$ while it does not affect significantly
the NCRIF. The reliability of such limit is confirmed by Ref. \cite{aoki}
where the magnitude of $\Delta[Q^{-1}]$ was shown to be sensibly affected
by changing the set-up frequency while the NCRIF was essentially
untouched.

We can see that the finite GB size BKT theory can naturally
account not only for the broad NCRIF transition but also for
the experimental height and broadness of the $\Delta[Q^{-1}]$ factor.
This latter point is not surprising since in the finite size
BKT framework, the broadness of the $n_s$ transition and of the
$\Delta[Q^{-1}]$ peak are related.
Although our theory can account for the difference of the signals
in a simple manner, the long tail of $n_s$ for 
$T > T_{\rm BKT}$ in the TOF sample
remains unexplained.
As mentioned above, however,
it should be noted that this feature of the NCRIF
has no counterpart in $\Delta[Q^{-1}]$,
which is sharply located at $T \approx T_{\rm BKT}$ \cite{chanHetre}.
This suggests that the long tail in $n_s(T)$
might be due in this sample to different physics
not related to the superfluid GB transition.

\begin{figure}[t]
\includegraphics[width=3.in]{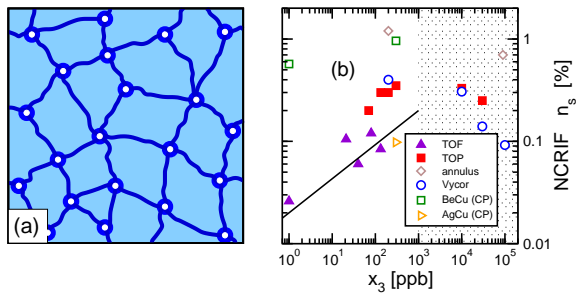}
\caption{(Color online) (a) schematic sketch of
GB structure of $^4$He samples in the presence
of $^3$He impurities.
Light blue areas represent solid $^4$He,
white spots $^3$He impurities and dark blue regions premelted $^4$He
at the GBs and around $^3$He impurities. (b) $n_s$ {\em vs.} $x_3$
plot of different samples (symbols) 
taken from Ref. \cite{chanHetre}
compared with our model prediction [Eq. (\ref{loglog})]
(solid line), which is expected to fail in the gray region
where $R \gtrsim 2-3 a_0$ and the mean-field theory breaks down.
}
\label{con}
\end{figure}

After assessed the robustness of our analysis
in explaining the NCRI phenomenology as function of $x_3$ and
growth conditions,
in the last part of this Letter we address
the origin itself of such $x_3$-dependence.
In particular we propose that the main effect
of $^3$He impurities is to provide an intrinsic
maximum length scale for the growth of grains, favoring thus the 
presence of grain boundaries, and hence to sustain a superfluid NCRI
on the GB surfaces.
It should be clear, on the other hand, that additional sources
of disorder, favoring the GB formation, can be present, as shown
by the different phenomenology for different growth conditions
and by the annealing dependence \cite{ann,RR}.
We assume for the moment that the only source of GB
is the presence of $^3$He impurities. This assumption is probably
valid for the TOF samples, which show the smallest NCRIFs for similar
$x_3$, and it is corroborated by the sharp drops of $n_s$ at $T_{\rm BKT}$,
suggesting quite large grains.
We consider, for dilute $^3$He concentrations $x_3 \lesssim 1$ ppm,
a uniform
distribution of the $^3$He impurities within the sample,
with an average distance $d_{^3{\rm He}}$ between $^3$He atoms
$d_{^3{\rm He}} \approx x_3^{-1/3}$.
In Ref. \cite{SM} it was shown that $^3$He impurities,
due to the stronger quantum fluctuations of zero point motion,
induce a local melting of the host $^4$He even at low temperature
much smaller than the bulk $^4$He melting.
The presence of local liquid spots,
which we remind survive also at virtually zero temperature,
around the $^3$He impurities
poses a strong constraint on the growth of the grain size
in the freezing process.
The average distance $d_{^3{\rm He}}$ provides thus
the maximum length scale for the growth of grain size and,
in the last analysis, an average value of the grain diameter
$2R_0 \simeq d_{^3{\rm He}}$
(see Fig. \ref{con}a for a sketched picture).
This simple estimate gives for instance $R_0 \simeq 1600$ \AA\
for $x_3=1$ ppb and $R_0 \simeq 313$ \AA\ for $x_3=133$ ppb, 2-3 times
larger than the actual fit estimates.
Taking into account that these figures
are purely indicative since
other sources of disorder are always present
further limiting the size of the grains, such estimates are not bad
and provide the order of magnitude of the grain size.
The most convincing probe of such picture is the power-law dependence
of the behavior of $n_s$ as function of $x_3$, $n_s \approx x_3^{1/3}$.
Assuming that a $p_s$ fraction of the thin liquid film of
thickness $d$ at the GBs undergoes
a superfluid BKT transition, and approximating the liquid $^4$He density
with the solid one, we have an areal SF density at the GBs
$\sigma_s=p_sd\rho$, and, from Eq. (\ref{MCMI}),
\begin{equation}
n_s\approx 6p_sd/d_{^3{\rm He}} \approx 6p_sd x_3^{1/3}/a,
\label{loglog}
\end{equation}
where we have assumed $d\ll R_0$ and
$R_0 \approx d_{^3{\rm He}}/2$.
where
$a=3.2$ \AA\ is the solid He-He distance at the pressure
here considered.
Such power-law behavior, assuming for instance $d=1.2$ \AA\ and
$p_s=0.01$,
is shown in Fig. \ref{con}b in qualitative agreement
with the experimental data for the TOF samples and the
lowest $x_3$ TOP samples \cite{chanHetre} where
other sources of disorder are thought to be small.
Note that the unknown quantities $p_s$, $d$, in the
log-log plot of Fig. \ref{con}b, determine only
the vertical off-shift of the $\log n_s - \log x_3$ behavior but not its
slope which is uniquely
determined by geometrical considerations.
In CP samples
on the other hand additional limiting mechanisms
on the grain size are probably operative concealing the $^3$He effects.
This is in agreement indeed with the larger values of $n_s$ and with the
broader drops of NCRIF \cite{aoki,chanHetre}.
The validity of this analysis is in addition limited by the
mean-field character of our approach.
For $x_3 \gtrsim 10^3$ ppm the average size of the grains
is predicted to be $R_0 \lesssim 150$ \AA, only 2-3 times
larger than the vortex size $a_0=50$ \AA.
In this regime the reliability of the mean-field analysis
is strongly questionable.

In conclusion, in this Letter we showed that the complex
NCRI phenomenology is fully compatible with a superfluid
Berezinski-Kosterlitz-Thouless
transition induced in the thin premelted liquid film at the grain boundaries.
Note by the way that the two dimensional BKT 
transition is characterized by 
the lack of specific heat anomalies,
in agreement with the heat-capacity
measurements in solid $^4$He \cite{day1,day2,todo}.
Both the temperature and $x_3$ dependence are shown to be ascribable
to finite grain size effects. We propose also a simple picture where
$^3$He impurities are directly related to the maximum size of the grains
and we predict a simple scaling relation between the NCRIF $n_s$
and the $^3$He impurity concentration.

We acknowledge useful discussions with J. Beamish, H. Alles, A.V. Balatsky.
M. Holzmann, S. Balibar and F. Caupin.

\end{document}